\documentclass[twocolumn,showpacs,amsmath,amssymb,prl]{revtex4}

\usepackage{graphicx}
\usepackage{dcolumn}
\usepackage{bm}

\begin{document}


\title{Electromagnetic Fluctuations during Fast Reconnection in a Laboratory Plasma}

\author{Hantao Ji}
\author{Stephen Terry}
\author{Masaaki Yamada}
\author{Russell Kulsrud}
\author{Aleksey Kuritsyn}
\author{Yang Ren}
\affiliation{Princeton Plasma Physics Laboratory, Princeton University,
	P.O. Box 451, Princeton, NJ 08543}

\date{\today}

\begin{abstract}

Clear evidence for a positive correlation is established 
between the magnitude of magnetic fluctuations 
in the lower-hybrid frequency range and enhancement of 
reconnection rates in a well-controlled laboratory plasma. 
The fluctuations belong to 
the right-hand polarized whistler wave branch, propagating 
obliquely to the reconnecting magnetic field, with a phase velocity 
comparable to the relative drift velocity between electrons and ions. 
The short coherence length and large variation along the propagation 
direction indicate their strongly nonlinear nature in three dimensions.

\end{abstract}


\maketitle

Magnetic reconnection~\citep[e.g.][]{biskampbook} plays an important role in 
determining the evolution of magnetic topology in relaxation processes in 
high-temperature laboratory plasmas, magnetospheric substorms, solar flares, 
and more distant astrophysical plasmas. Often, magnetic reconnection 
is invoked to explain the observed rapid release of magnetic energy in 
these highly conducting plasmas. A central question of magnetic reconnection 
concerns why the observed reconnection rates are much faster than predictions 
by the Sweet-Parker model~\cite{sweet58,parker57} based on 
magnetohydrodynamics (MHD) with the classical Spitzer resistivity.
In this two-dimensional (2D) model, the infinitely small resistivity 
causes magnetic field to dissipate 
only in very thin current sheets, which impede the outflow of mass
leading to significantly slow reconnection rates. The subsequently 
proposed Petschek model~\cite{petschek64} is based on standing slow 
shocks to open up the outflow channel allowing larger mass flows thus 
the faster reconnection rates. However, it has been shown 
later~\cite{biskamp86,uzdensky00} that the Petschek solution is 
not compatible with uniform or smooth resistivity profiles. 
On the other hand, perhaps not surprisingly, the plasma resistivity
can be enhanced due to microinstabilities which are often
active only in the reconnection region
where plenty of free energy exists in the form of a large relative drift between 
ions and electrons and large inhomogeneities in 
pressure and magnetic field. 
This anomalous resistivity not only can broaden the current 
sheet to increase the mass flow and the reconnection rate
in the context of the Sweet-Parker model~\cite{kulsrud01}
but also its localization is able to open up the outflow channel 
for the fast reconnection~\cite{ugai77,kulsrud01,biskamp01}. 
Alternatively, a recent theory~\cite{birn01} attempts to explain 
fast reconnection rates based on non-dissipative terms, 
notably the Hall term, in the generalized Ohm's law in a 2D and laminar fashion.

The underlying microinstabilities~\cite{biskampbook} for the resistivity enhancement
have been considered to be predominantly electrostatic in nature
due to their effectiveness in wave-particle interactions.
The primary candidate is the Lower-Hybrid Drift Instability 
(LHDI)~\cite{krall71}, which has been frequently observed in space 
plasmas~\cite{gurnett76,cattell86}. Recently this instability has been
studied in direct relation with reconnection in laboratory~\cite{carter02a}
and in space~\cite{shinohara98,bale03}. The main conclusion of these studies
is that the LHDI is active only at the low-$\beta$ edge region of current sheet, 
but {\it not} at the high-$\beta$ central region, where the resistivity needs to 
be enhanced for fast reconnection. This conclusion is 
consistent with an earlier theory~\cite{davidson77} which showed 
stabilization of LHDI by finite plasma $\beta$, and also with more recent 
numerical simulations~\cite{horiuchi99,lapenta02}.

By contrast, much less attention has been paid to electromagnetic fluctuations
and their relation with reconnection, although they have been also regularly 
observed in space~\cite{gurnett76}. In this Letter, we report the first clear experimental 
evidence for electromagnetic fluctuations in the lower-hybrid frequency range
during fast reconnection in a well-controlled laboratory plasma, Magnetic Reconnection 
Experiment (MRX)~\cite{yamada97b}. The observed waves are identified 
as right-hand polarized whistler waves propagating obliquely 
to the magnetic field. Earlier laboratory experiments~\cite{stenzelgekelman6} 
on reconnection have indicated evidence for high-frequency electromagnetic 
fluctuations in the electron MHD regime where only electrons are 
magnetized, but their roles in the reconnection process were unclear. 
On the other hand, most plasmas of interest for the reconnection 
problem are well into the MHD regime where ions are also magnetized at least outside 
of the reconnection region, as in MRX. In one case, 
magnetic fluctuations have been interpreted as byproducts of 
LHDI in space plasmas~\cite{shinohara98}.

\begin{figure}[t]
\centerline{
\includegraphics[width=2.5truein]{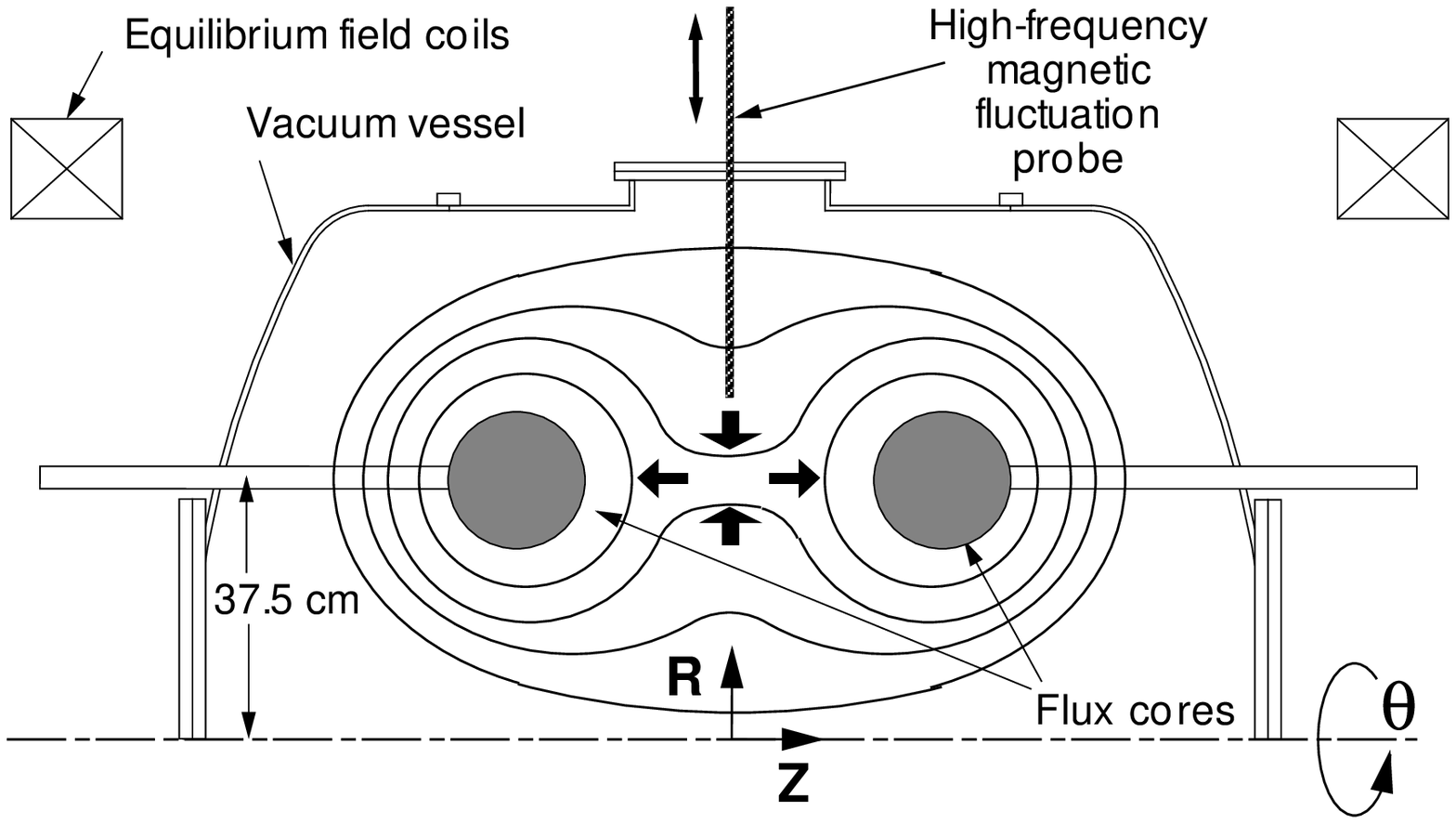}}
\caption{Experimental setup in MRX for magnetic reconnection
which is induced by reducing coil currents in the \lq\lq flux 
cores\rq\rq. Oppositely directed magnetic field lines is driven towards 
each other to form a current sheet flowing 
in the $\theta$ direction. Inflow and outflow 
are in the radial (R) and axial (Z) directions, respectively.} 
\label{figure:setup}
\end{figure}

In the MRX, magnetic reconnection is driven by reducing currents in two flux cores 
whose toroidally symmetric shape ensures the global 2D geometry, as 
illustrated in Fig.\ref{figure:setup}. Oppositely directed field
lines in the $Z$ direction are pulled together radially, forming a current sheet
flowing along the $\theta$ direction. All essential parameters needed 
for characterizing reconnection are measured by an extensive set 
of diagnostics~\cite{yamada97b}. In a previous quantitative study~\cite{ji98}, 
it was shown that the observed reconnection rates can be explained 
by a modified Sweet-Parker model including an effective resistivity, 
determined experimentally by $\eta^* \equiv E_\theta/j_\theta$ 
where $E_\theta$ and $j_\theta$ are toroidal reconnecting electric field 
and current density, respectively. In highly collisional plasmas,
$\eta^*$ is very close to the classical Spitzer perpendicular 
resistivity $\eta_{\text{Spitzer}}$~\cite{trintchouk03} while $\eta^*
\gg \eta_{\text{Sptizer}}$ when the collisionality is reduced~\cite{ji98}, in 
correlation with strong nonclassical ion heating~\cite{hsu00}. 
High-frequency fluctuations have been measured in MRX in order to 
explore the possibility of the resistivity enhancement due to
microinstabilities in low collisionality regimes. 
Although the LHDI was identified at the low-$\beta$ edge region of 
the current sheet, electrostatic fluctuations did not correlate well 
with the reconnection process in their temporal and spatial behavior 
as well as their collisionality dependence, leading to a conclusion 
that the electrostatic fluctuations do not play an essential role 
in the fast reconnection in MRX~\cite{carter02a,carter02b}.

\begin{figure}[b]
\centerline{
\includegraphics[width=3.0truein]{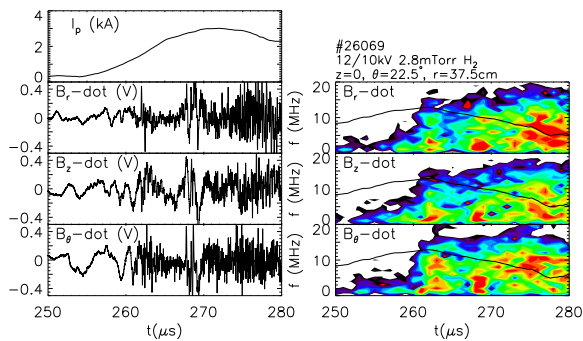}}
\caption{Traces of typical raw signals during reconnection represented 
by plasma current (top left panel). Spectrograms of each signal are 
also shown on the right panels where fluctuation powers are color-coded 
(decreasing power by order of red, yellow, green, blue and white)
in the time-frequency domain. The black lines indicate local $f_{\text{LH}}$.}
\label{figure:example}
\end{figure}

The experimental results reported here focus on electromagnetic 
fluctuations, which do correlate well with fast reconnection. 
The main diagnostics used are based on four small magnetic pickup coils 
mounted inside an electrostatically shielded glass tube. 
All three components of magnetic field are measured at almost the same location.
The probe outputs are fed into a miniature circuit board, which houses
four amplifiers embedded in the probe shaft near the tip in order to
provide noise immunity and impedance matching. The integrated bandwidth
is up to 30 MHz.
\begin{figure}[b]
\centerline{
\includegraphics[width=2.2truein]{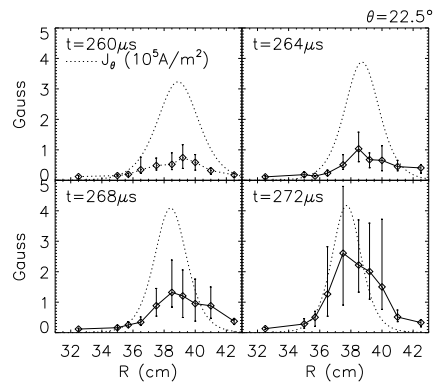}}
\caption{Radial profile of total amplitude of magnetic fluctuations 
with $f\ge 1$MHz at four time points. The current density profiles
(dotted lines) are also shown for reference.}
\label{figure:profile}
\end{figure}

Typical raw signals are shown in Fig.\ref{figure:example} during
a single discharge when reconnection is induced.
High-frequency fluctuations appear in all three components of 
magnetic field
right after $t=260\mu$s when the current sheet is formed,
and persist as long as the reconnection goes on.
Spectrograms, which display the fluctuation power in the time-frequency
domain, are shown in the right panels of 
Fig.\ref{figure:example}.
It is seen that discrete peaks exist in the lower-hybrid 
frequency range and they tend to also vary accordingly with the changes 
in the local lower-hybrid frequency $f_{\text{LH}} \equiv 
\sqrt{f_{\text{ce}}f_{\text{ci}}}$, shown as black lines~\footnote{ 
The local magnetic field (and $f_{\text{LH}}$), 
remains nonzero at the current sheet center due to a residual 
toroidal field generated during the plasma formation process~\cite{yamada97b}.}.
The spatial profile of the total
fluctuation amplitude $|\widetilde B|$ is plotted in 
Fig.\ref{figure:profile}. The fluctuations have large amplitudes consistently 
near the current sheet center with peak $|\widetilde 
B|/B_{\text{up}}$ up to 5\%, 
where $B_{\text{up}}$ is the upstream reconnecting magnetic field. 
It should be 
noted that both temporal behavior and spatial behavior of the magnetic 
fluctuations are in sharp contrast with the 
electrostatic fluctuations~\cite{carter02a,carter02b}.

\begin{figure}[t]
\centerline{
\includegraphics[width=2.8truein]{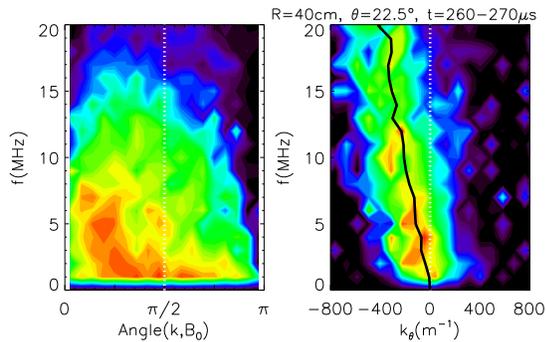}}
\caption{Power spectra (color-coded as in Fig.\ref{figure:example})
for $\widetilde B$ in the domain of frequency and the angle between 
dominant $\bm{k}$ and $\bm{B}_0$ (left) and for $\widetilde B_Z$
in the domain of frequency and $k_\theta$ (right) at $R=40$cm.}
\label{figure:spectra}
\end{figure}

In order to identify the observed electromagnetic waves,
it is crucial to measure their propagation characteristics.
Two techniques have been employed for this purpose. The first one
is called the hodogram technique~\cite{urrutia00}, which is based
on the tip trajectories of a fluctuating magnetic field vector measured at
a single point in space.  For a given frequency,
the condition $\nabla \cdot \widetilde {\bm{B}} =0$ can be 
translated to $\bm{k} \cdot \widetilde{\bm{B}} =0$ where $\bm{k}$ 
is the dominant wavenumber vector, the direction of 
$\bm{k}$ can be determined by the right-hand rule 
$\widetilde{\bm{B}}(t_0) \times \widetilde {\bm{B}}(t_0+\delta t)$
when $f \gg f_{\text{ci}}$.
Power spectra of the fluctuating magnetic field can be 
constructed in the domain of frequency and the angle between $\bm{k}$ and 
the background $\bm{B}_0$, as shown in Fig.\ref{figure:spectra}(left). 
It is seen that 
in the low frequencies, $\bm{k}$ has a rather large angle 
($\sim 60^\circ$) to $\bm{B}_0$ with a broad spread while in higher 
frequencies, $\bm{k}$ has a rather small angle ($\sim 30^\circ$)
to $\bm{B}_0$ with a narrow spread. In addition, since the 
$\bm{k}$ vector has only a small radial component (not shown), 
$\bm{k}$ remains in the $Z-\theta$ plane. Therefore, 
it is concluded that the observed magnetic fluctuations are
right-hand polarized whistler-like waves propagating obliquely  
to the magnetic field while staying within the 
current sheet.

While the direction of $\bm{k}$ is determined by the hodogram technique,
its magnitude or the wavelength is determined by a second
technique using the phase shift measured between two spatial points.
Two glass tubes, each containing a pickup coil, 
are mounted to a probe shaft with a distance 
of 7 mm between their axes. Each signal is fed to a single-channel version 
of the miniature in-shaft amplifier~\cite{carter02a}. 
The phase shift along the $\theta$ or $Z$ direction
is measured to construct power spectra in the 
domain of frequency and $k_\theta$
[Fig.\ref{figure:spectra}(right)] or $k_Z$ (not shown).
It is found that the magnetic fluctuations propagate mainly in 
the $-\theta$ direction, along which electrons drift, but not in the $Z$ 
direction. The phase velocity, $V_{\text{ph}}=(3.4\pm 0.8)\times 10^5$m/s
calculated from the slope of the black line in Fig.\ref{figure:spectra}(right), 
is reasonably consistent with the relative drift velocity $V_{\text{d}}
\equiv j_\theta/en=(2.5\pm 0.9)\times 10^5$m/s, where $n$ is the plasma density.

The measurements of the propagation characteristics described above are 
made at the outer edge of the current sheet ($R$=40cm) where the 
fluctuations have only moderate amplitudes (see 
Fig.\ref{figure:profile}) and stay relatively coherent within the probe 
separation. At the current sheet center where the 
fluctuation amplitude peaks, the measurements of $\bm{k}$ have not been so
successful due to extremely short coherence lengths.
The coherence 
$\gamma\equiv |\widetilde B_1\widetilde B_2^*|/(|\widetilde B_1||\widetilde B_2|)$ 
between signals $\widetilde B_1$ and $\widetilde B_2$ rapidly decreases to
the noise level when the separation is larger than 0.5$-$1.5 cm, 
suggesting their strongly nonlinear nature. 
(For comparison, the wavelength measured at $R=$40cm in 
the same $\theta$ direction is $\sim 5$cm for $f$=10MHz.)
Furthermore, it is found that the fluctuation amplitude 
varies substantially along the $\theta$ (current) direction, breaking 
the 2D axisymmetry of the current sheet and consequently the reconnection 
process. Further detailed measurements indicate that the fluctuation amplitude
tends to correlate with the {\it local} $V_{\text{d}}$.

\begin{figure}[b]
\centerline{
\includegraphics[width=2.8truein]{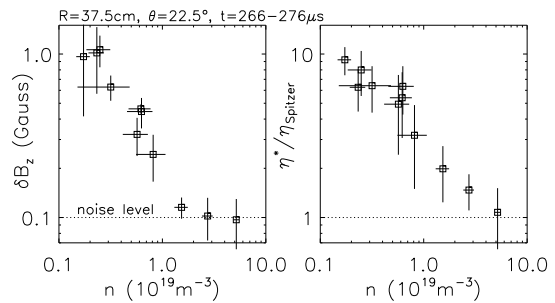}}
\caption{Density dependence of $\widetilde B_Z$ amplitude 
at the current sheet center and resistivity enhancement.}
\label{figure:eta_B_n}
\end{figure}

\begin{figure}[t]
\centerline{
\includegraphics[width=1.8truein]{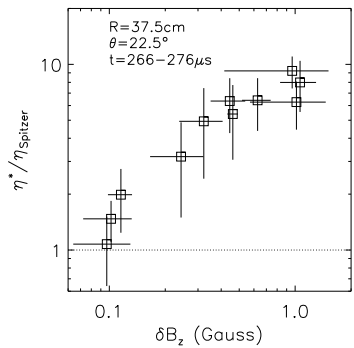}}
\caption{Resistivity enhancement versus fluctuation amplitude of 
$\widetilde B_Z$ at the current sheet center.}
\label{figure:eta}
\end{figure}

The next important question concerns how the observed magnetic 
fluctuations are related to the resistivity enhancement, and thus the 
fast reconnection process in MRX. It is consistently found that 
the fluctuation amplitudes are sensitive to plasma density
or equivalently the collisionality. For example,
when the density is reduced from $\sim 5\times 10^{19}{\text{m}}^{-3}$
to $\sim 2\times 10^{18}{\text{m}}^{-3}$, $|\widetilde B_Z|$
increases from 0.1 G (close to the noise level for the 
measurements) to $\sim 1$ G, as shown in Fig.\ref{figure:eta_B_n}. 
Since the resistivity enhancement also strongly depends on the plasma 
collisionality~\cite{ji98} (Fig.\ref{figure:eta_B_n}),
a clear positive correlation between magnetic fluctuations and 
resistivity enhancement is established, as shown in Fig.\ref{figure:eta}
\footnote{The resistivity is calculated by assuming toroidal symmetry
of the current sheet~\cite{ji98}, which does not necessarily hold true in 
the discharges used for this study. However, the averaged resistivity
obtained by a scan over 20 toroidal angles is reasonably consistent with 
values based on the axisymmetry assumption provided enough numbers of 
discharges are used for averaging.}.

Given the experimental data described above, it is clear that the observed 
electromagnetic 
waves are caused by microinstabilities driven by free energy stored in the 
reconnecting current sheet, in the form of large $V_{\text{d}}$ and/or large
inhomogeneities. 
Since essentially no electrostatic fluctuations are detected concurrently, 
the observed magnetic fluctuations cannot be simply byproducts of LHDI, which is 
driven by inhomogeneities, propagating primarily perpendicular to the 
field. In addition, magnetic fluctuations due to 
LHDI are dominated by their component along $\bm{B}_0$ while 
experimentally all three components have roughly the same amplitude 
(see Fig.\ref{figure:example}). The waves also cannot be 
described as whistler waves driven by electron temperature 
anisotropy~\cite{kennel66}, which propagate mainly along $\bm{B}_0$.

Many observed key features of the magnetic fluctuations, however, are 
consistent with the so-called Modified Two-Stream Instability (MTSI) 
in the high-$\beta$ limit~\cite{ross70,krall71,mcbride72,wu83,basu92}.
In contrast to LHDI, MTSI is driven only by a large relative drift
across the magnetic field in {\it homogeneous} plasmas. 
In the low-$\beta$ limit, both LHDI and MTSI behave similarly~\cite{silveira02}.
When $\beta$ is large ($\agt 1$), LHDI is stabilized~\cite{davidson77} while 
MTSI remains unstable but the Alfv\'en speed $V_{\text{A}}$ replaces 
the ion sound (or thermal) speed as the critical speed for $V_{\text{d}}$.
The resultant waves are largely electromagnetic and right-hand polarized
whistler-like waves propagating obliquely to the field with
$V_{\text{ph}} \sim V_{\text{d}}$~\cite{ross70}.
Under certain conditions similar to those at the current sheet in MRX,
such as $\beta \agt 1$ and $V_{\text{d}}/V_{\text{A}} \agt 5$, the waves 
are unstable only at certain propagation angles to the 
field~\cite{wu83,basu92}. However, the discrete peaks in the frequency
spectra (Fig.\ref{figure:example}) cannot be explained by these 
theories, which were based on the slab geometry. 
Global eigenmode calculations taking into account the boundary 
conditions, as have been done for LHDI~\cite{yoon02,daughton02},
may explain the discrete frequency peaks.

Quantitative estimates of resistivity enhancement due to these fully nonlinear
waves as described above are not straightforward. Nonetheless,
examinations of each term in the generalized Ohm's law reveal
that the turbulent Hall term due to the magnetic fluctuations 
could be sufficiently large to balance the reconnecting electric field if 
the measured coherence lengths are used as
the perpendicular scale length. Other candidates, such as 
off-diagonal terms in the pressure tensor, can also possibly provide
required effective friction between electrons and ions as a result of
wave-particle interactions. 

In summary, a detailed experimental study in MRX has established, for 
the first time, a clear and positive correlation between magnetic 
fluctuations in the lower-hybrid frequency range and fast reconnection 
in the low-collisionality regimes. The waves have been identified as
right-hand polarized whistler waves, propagating obliquely to the 
reconnecting field, with a phase velocity comparable to the relative 
drift velocity. These waves are consistent with the modified two-stream
instability driven by large drift speeds compared to the Alfv\'en
speed in high-$\beta$ plasmas. The short coherence length and large 
variation along the propagation direction indicate their strongly
nonlinear nature. Quantitative understanding of the effects of these 
waves on the resistivity enhancement, and thus reconnection rates, requires
further experimental efforts to directly measure relevant terms 
in the generalized Ohm's law. Theories and numerical studies 
of linear and nonlinear characteristics of current-driven microinstabilities
using proper models and boundary conditions in 3D should provide
much-needed physical insight.

The authors are grateful to D. Cylinder and R. Cutler for their
excellent technical support, J. Whitney, K. Shen, and J. Carter
for their contributions through a summer program, and T. Carter for useful 
discussions. This work was jointly supported by DOE, NASA, and NSF.


\end{document}